\documentclass[openacc]{rstransa}

\begin{document}

\title{
Nonlinear dynamics determines the thermodynamic instability of condensed matter in vacuo
}
\author{Julyan H. E. Cartwright$^{1,2}$}

\address{$^{1}$Instituto Andaluz de Ciencias de la Tierra, CSIC--Universidad de Granada, E-18100 Armilla, Granada, Spain \\
$^{2}$Instituto Carlos I de F\'{\i}sica Te\'orica y Computacional, Universidad de Granada, E-18071 Granada, Spain \\
}

\subject{nonlinear dynamics, statistical physics, thermodynamics}


\corres{Julyan Cartwright\\
\email{julyan.cartwright@csic.es}}

\begin{abstract}
Condensed matter is thermodynamically unstable in a vacuum. That is what thermodynamics tells us through the relation showing that condensed matter at temperatures above absolute zero always has non-zero vapour pressure.  This instability implies that at low temperatures energy must not be distributed equally among atoms in the crystal lattice but must be concentrated. In dynamical systems such concentrations of energy  in localized excitations are well known in the form of discrete breathers, solitons, and related nonlinear phenomena. It follows that to satisfy thermodynamics such localized excitations must exist in systems of condensed matter at arbitrarily low temperature and as such the nonlinear dynamics of condensed matter is crucial for its thermodynamics.
\end{abstract}

\begin{fmtext}

\section{Introduction}

How thermodynamics, the limit of large numbers of degrees of freedom and their statistical properties that lead to laws that are extremely well observed in nature, and nonlinear dynamics of systems of small numbers of degrees of freedom with their deterministic behaviour, fit together is an extremely pertinent and deep question. One place where this interface between nonlinear dynamics and thermodynamics is apparent is in the behaviour of condensed matter at low temperature.

\end{fmtext}

\maketitle

\section{Historical digression}

The roots of  nonlinear dynamics, statistical physics, and thermodynamics all lie in the 19th-century. In an 1859 letter to Stokes, preserved in Stokes' {\em Memoirs} \cite{larmor1907}, Maxwell wrote
\begin{quote}
``as I found myself able and willing to deduce the laws of motion of systems of particles acting on each other only by impact, I have done so as an exercise in mechanics. Now do you think there is any so complete a refutation of this theory of gases as would make it absurd to investigate it further so as to found arguments upon measurements of strictly `molecular' quantities before we know whether there be any molecules?''
\end{quote}
By 1873, he could affirm
\begin{quote}
``I think the most important effect of molecular science on our way of thinking will be that it forces on our attention the distinction between two kinds of knowledge, which we may call for convenience the Dynamical and Statistical.''
\end{quote}
and 
\begin{quote}
``if the molecular theory of the constitution of bodies is true, all our knowledge of matter is of the statistical kind.''
\end{quote}
At the same time, Maxwell began to see that, for systems composed of a large number of bodies, a sensitive dependence on the initial conditions might exist
\begin{quote}
``There are certain classes of phenomena ... in which a small error in the data only introduces a small error in the result. ... The course of events in these cases is stable. There are other classes of phenomena which are more complicated, and in which cases of instability may occur, the number of such cases increasing, in an exceedingly rapid manner, as the number of variables increases.''
\end{quote}
Thus Maxwell, at least, was already thinking of the complexities introduced into physics by both many-body systems and nonlinear systems. It is a great pity that the letters from Stokes to Maxwell seem to be lost. In his published work and even in his extant letters to people whom he does not know well, Stokes is so guarded as to give us almost no idea of his real thoughts on molecules, statistical physics, and so on. 

Stokes' editor Larmor \cite{larmor1907} asserted that 
\begin{quote}
``With Maxwell the scientific imagination was everything: Stokes carried caution to excess. Maxwell revelled in the construction and dissection of mental and material models and images of the activities of the molecules which form the basis of matter: Stokes' published investigations are mainly of the precise and formal kind, guided by the properties and symmetries of matter in bulk, in which the notion of a molecule need hardly enter.''
\end{quote}
But it has turned out, over the intervening century, that understanding both molecules and ``investigations ... of the precise and formal kind'' on the ``properties and symmetries of matter in bulk'' is required to make progress in many areas of physics. 

The history linking  continuum and discrete aspects of the question at hand is convoluted and interesting. In 1834 Russell observed solitary water waves \cite{russell1845}. Stokes proposed in 1847 that what Russell had observed were in fact periodic wave trains in a nonlinear system \cite{stokes1847}. Benjamin and Feir demonstrated in 1967 that these Stokes waves could become unstable \cite{benjamin1967}. Lake {\em et al} showed in 1977 that under certain circumstances the so-called Benjamin--Feir instability may lead to a Fermi--Pasta--Ulam--Tsingou-type recurrence \cite{lake1977}. Meanwhile an approximate theory for a solitary wave was given by Boussinesq in 1871 \cite{boussinesq1871} and Rayleigh in 1876 \cite{rayleigh1876}. In 1895, Korteweg and de Vries derived  a full nonlinear theory \cite{korteweg1895}. In 1965, Zabusky and Kruskal noted that  the Korteweg--de Vries equation, the integrable continuum limit of the discrete Fermi--Pasta--Ulam--Tsingou system, possesses soliton solutions \cite{zabusky1965}.

\section{The Fermi--Pasta--Ulam--Tsingou system}

It took until the middle of the 20th century, with the development of the computer, for a crucial numerical experiment linking molecules and bulk matter to be performed. Having the thermalization of a solid  in mind, in the 1950s Fermi, Pasta, Ulam, and Tsingou \cite{fpu,dauxois2008} examined the problem of how a chain of atoms linked together would vibrate. With a numerical study of the model
$$\ddot x_n = V'(u_{n+1}) - V'(u_n);  \quad u_n = x_n - x_{n-1}, $$
where the potential $V(u_n)$ could have  a quadratic ($\alpha\neq 0$) or a cubic  ($\beta\neq 0$) nonlinearity 
$$V(u_n) = \frac{1}{2} u_n^2 + \frac{\alpha}{3} u_n^3 + \frac{\beta}{4} u_n^4, $$
using one of the first electronic computers they found to their surprise that nonlinearity in the bonds between atoms did not lead to what they expected to confirm, to equipartition of energy along the chain, but instead to successive recurrences of states where energy was concentrated in one oscillator.
What Fermi, Pasta, Ulam and Tsingou had unexpectedly found are localized excitations of the lattice. The quest for the  understanding of these localized excitations has led to many decades of work in nonlinear dynamics \cite{berman2005}.

Let us make some thought experiments to consider whether there could exist a threshold in temperature below which thermal agitation is insufficient to break bonds in a crystal lattice and so be unable to liberate an atom of the solid. If such a threshold existed, the vapour pressure would be zero below a given cutoff temperature. 
First, if one has a classical ideal gas with no interatomic interactions --- solid only at 0 K --- then, because there are no interatomic interactions, the argument that no such cutoff exists clearly holds. Now let us add bonds between atoms that must be broken to disassociate the material. If it holds that energy is distributed equally in a form of equipartition, then these bonds will be broken simultaneously at the sublimation temperature. If, on the other hand, there is a distribution of energies with an arbitrarily long tail, such as a Maxwell--Boltzmann distribution --- which, recall,  is the distribution for an ideal gas (see a statistical mechanics textbook, e.g., \cite{tolman}) --- then there will always be the possibility of an atom having sufficient energy to break its bonds and to sublimate,  so again agreeing with the idea that no such cutoff should exist.  But what is the distribution of energies in a solid? This is where the Fermi--Pasta--Ulam--Tsingou results   have something to say. They demonstrate that in such a nonlinear dynamical system energy is not distributed equally in space and time.

\begin{figure}
\centering\includegraphics[width=0.5\columnwidth]{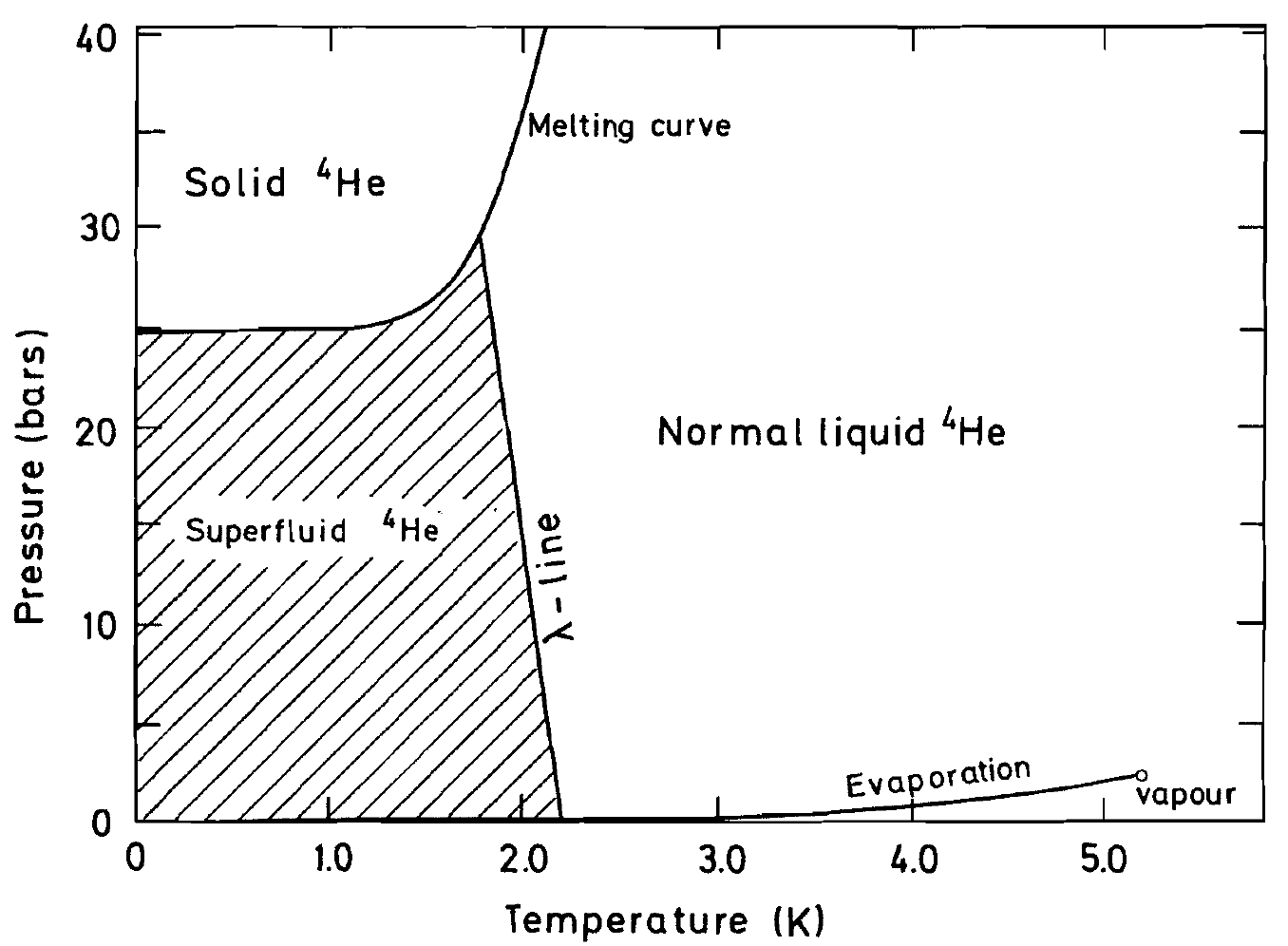}
\caption{How the sublimation curve originates. Low temperature phase diagram of helium 4, $^4$He \cite{vollhardt}, showing at the bottom of the figure the  curve approaching the origin of zero temperature and pressure tangentially to the temperature axis. Below the curve, the gas is the stable phase, while above it  condensed matter is stable. There are complexities in this phase diagram that are specific to $^4$He, which remains liquid, rather than solid, even at the lowest temperatures above the curve, and moreover has a superfluid phase as well as a normal liquid phase. For this reason the curve, terminating in the critical point beyond which liquid and gas are unified as a supercritical fluid, is labelled \emph{evaporation} rather than \emph{sublimation}, but its behaviour is nonetheless universal. 
}
\label{fig2}
\end{figure}

\section{The thermodynamic instability of condensed matter in vacuo}

\begin{figure}
\centering\includegraphics[width=0.5\columnwidth]{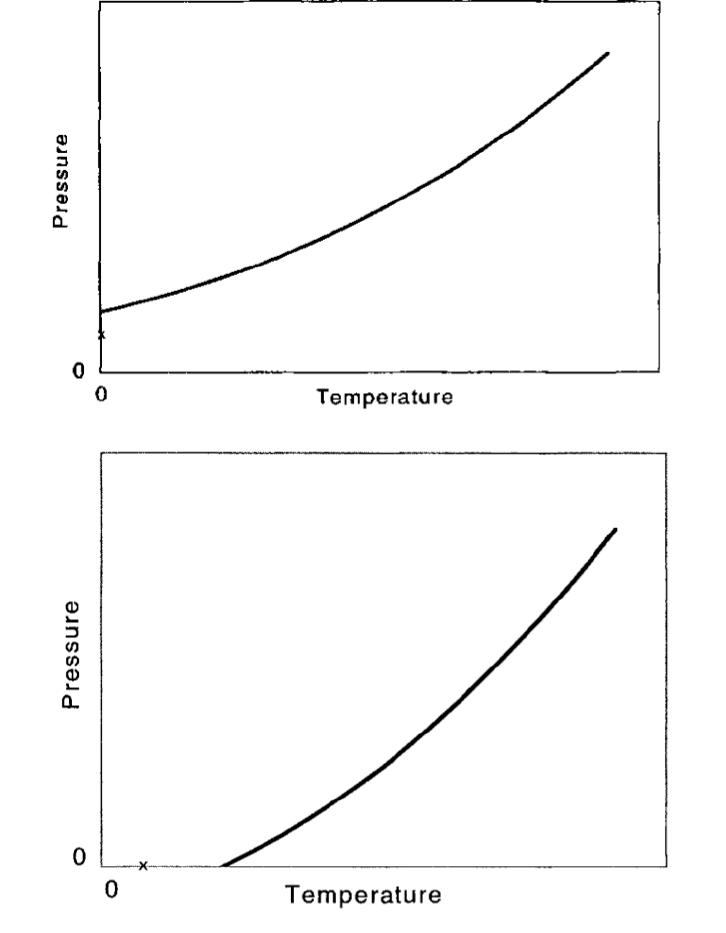}
\caption{How the sublimation curve does not originate. Above, if the sublimation curve should end at a positive point on the pressure axis, then a gas would be thermodynamically stable at zero temperature and positive pressure; below, if the sublimation curve should end at a positive point on the temperature axis, then a solid would be  thermodynamically stable at zero pressure and positive temperature \cite{Diu2002}.}
\label{fig1}
\end{figure}

The sublimation curve represents the conditions under which the solid condensed phase of a substance and its vapour phase are in equilibrium; on one side of the curve, the solid is stable, while on the other, the gas is the stable phase.
Let us recapitulate the thermodynamical argument that the sublimation curve ends at the origin of zero pressure and temperature, with a tangential separation from the temperature axis, as seen in Fig.~\ref{fig2}.
To begin with, one can appreciate that if a gaseous phase were thermodynamically stable at a positive pressure at zero temperature, as in Fig.~\ref{fig1}a, then one would need to posit fluctuations at zero energy to avoid the gas condensing. (Zero point energy does exist in a quantum system; liquid helium retains kinetic energy and does not freeze regardless of temperature owing to zero-point energy; nonetheless it is still a condensed phase, not a gas.)
If, on the other hand, a a condensed phase were thermodynamically stable at a positive temperature at zero pressure,  as in Fig.~\ref{fig1}b, then it would imply that there exists a threshold in energy below which thermal agitation would be too weak to allow atoms to escape from their neighbours in the condensed phase to become a gas.
Now, the thermodynamic equilibrium between a solid and its vapour is given by equating their chemical potentials \cite{baierlein2001}
$$\mu_s=\mu_g.$$
A standard result is that the chemical potential of a vapour at low pressure $p$ and temperature $T$, a classical ideal gas, is
$$\mu_g= -RT  \ln (RT/(p\lambda^3))$$ where $R$ is the gas constant and $\lambda  = h/(2\pi m k T)^{1/2}$ is the thermal de Broglie wavelength ($h$ is Planck's constant, $m$ the mass of the particle, $k$, Boltzmann's constant).
 Thus the equilibrium vapour pressure $p_e$ is 
 $$p_e = (2\pi m/h^2)^{3/2} (kT)^{5/2} \exp (\mu_s /kT)$$
 in terms of the chemical potential of the solid.
Then to calculate the vapour pressure, we need a model of the chemical potential of the solid, $\mu_s$. 
If we take the common quasi-harmonic model \cite{dove,venables}, 
$$\mu_s = U_0 + \langle 3h\nu/2 \rangle + 3kT \langle\ln(1-\exp(-hn/(kT)))\rangle$$
(angle brackets denote averaging; $\nu$, frequency; and $U_0$ is the energy per particle in the solid relative to the vapour phase),
then we  may approximately obtain 
$$\exp (\mu_s /kT) = (h\nu/kT)^3 \exp (-L_0/kT),$$
where $L_0$ is the sublimation energy at absolute zero,  so that 
 $$p_e = (2\pi m \nu^2)^{3/2} (kT)^{-1/2} \exp (-L_0 /kT);$$
 i.e., a tangential approach with zero pressure at zero temperature.
 The foregoing is an illustrative derivation with a typical model; Diu et al \cite{Diu2002} derive this relation showing that the sublimation curve ends at the origin of zero temperature and pressure in the most general way\footnote{Diu et al \cite{Diu2002} note that their self-consistent derivation is ``based on two foundation stones, essentially: we treated the vapour phase as a classical ideal gas, and took its specific heat at constant volume as independent of $T$ (equal to $3R/2$). The first assertion is proved in [their] section 4. The second one is readily fulfilled by all monoatomic gases (noble gases, metallic vapours, ...). More complex molecules can rotate, and their atoms vibrate. The vibrational degrees of freedom are already frozen below several hundreds of kelvins, thus raising no problem. The limiting phenomenon is then rotation. Its freezing point largely varies along with the substance: 85~K for hydrogen, 0.35~K for chlorine, 0.56~K for carbon dioxide, ... As far as we know, no real sublimation curve has truly been followed down to such temperatures.''} in which they show the self consistency of the steps involved.

Thus for the thermodynamics to hold,  above absolute zero there always has to be a nonzero vapour pressure.
The profound implication of this result is that, in terms of thermodynamics, all condensed matter is unstable in vacuo.  Its apparent stability is down to aspects of condensed matter beyond thermodynamics; to kinetics --- to the slowness of this evaporation process --- and to gravitation for larger amounts of condensed matter where self-gravitation becomes important. Out of equilibrium the temperature will decrease as more energetic, hotter atoms escape from the solid, but since we are in the thermodynamic limit there is always a reservoir of energy in the solid. Also, since thermodynamics is concerned with large quantities of material one cannot make the argument of a total amount of energy in the material that could be posited to be less than the energy of the bonding.
In any case, a discrete amount of material will not have an infinitely long tail: there must be a cutoff, an atom with the greatest energy (at any given time), and if the cutoff is below the bond-breaking threshold, then the material is stable. But this lack of infinite scale breaks the thermodynamic suppositions and takes us to the behaviour of `small' (i.e., finite; 'small' may be very large but finite for a material with a low vapour pressure; metal or rock, say) numbers of atoms.
The Hertz--Knudsen--Langmuir equation \cite{kolasinski} describes the evaporation rate as 
$$ \alpha p_e/(2\pi m k T)^{1/2}$$ where $\alpha$ is a sticking coefficient between zero and one.
For the minuscule vapour pressures at low enough temperatures, real materials are simply metastable, but  with a lifetime for the evaporation of the condensed phase that can be arbitrarily long.

We should like to know the behaviour of real substances at low temperatures, for instance, in astronomy, where it is of interest to understand how condensed matter behaves in the cold ($\sim3$~K) of interstellar space. One of the most abundant molecules in the universe is water, and so  theoretical work has examined how its solid phase, ice \cite{bartels2012},  should sublime at  low temperatures and pressures \cite{feistel2007}. For example, one piece of work, interested in the possible behaviour of ice on the Moon in permanently shaded craters near the lunar poles, where a typical temperature might be 40~K, found that ``for temperatures below 70~K, the sublimation rate of an exposed ice surface is much less than one molecule of water vapor lost per square centimeter of surface per hour'' \cite{andreas2007}.
There is not a great deal of experimental confirmation of the behaviour of the sublimation curve approaching absolute zero.  At the lowest temperatures, close to absolute zero, there is work on helium \cite{vollhardt,angus}, as illustrated in figure~\ref{fig2} for helium 4. In fact, the behaviour of  the sublimation curve approaching the origin of zero temperature and pressure tangentially to the temperature axis, in agreement with the thermodynamic argument presented above, has been used to determine  temperatures
below 5~K from the saturated vapour pressure of liquid helium.

\section{Localized excitations: solitons and discrete breathers}

The initial work of Fermi, Pasta, Ulam and Tsingou looked at low-frequency (long wavelength) excitations of the chain of atoms. The responses of the system, the localized excitations in the form of  successive recurrences of states where energy was concentrated in one oscillator, were established to be solitary travelling waves, now termed solitons.
More recently, it is found that if instead of beginning with a low-frequency initial state, one sets initial conditions of  high-frequency (short wavelength), localized excitations that have an oscillating amplitude, called 
discrete breathers (also called intrinsic localized modes) \cite{flach1998,mackay2000,flach2008}  appear. This is true both in the Fermi--Pasta--Ulam--Tsingou system \cite{james2001,reigada2002,flach2005} and in other oscillator chain models \cite{velarde2016,cuevas2017}. With general initial conditions of the chain, both solitons and discrete breathers can be found and one may transform into the other  \cite{pistone2018}.
It should be noted that localization of excitations is to be expected in crystals with defects, or heterogeneities, in the lattice; this is known as Anderson localization \cite{anderson1958}. What was unexpected was finding such localization in a homogenous chain without defects. 

Discrete breathers are now being sought for in applications \cite{mackay2000,flach2008}, and are being found in crystals \cite{dmitriev2016}. For example,  work is ongoing regarding discrete breathers in graphene \cite{liu2013} and in clay and mica minerals \cite{archilla2006}. As well as classical breathers, their quantum analogues are being analysed \cite{mackay2000,ivic2006} and noted in experiments as on solid hydrogen \cite{gush1957}.
However, the rather intriguing link between the thermodynamics of sublimation and vapour pressure and the necessity of the existence of  localized excitations in general in crystals close to absolute zero does not appear to have been highlighted.

The existence of localized excitations down to all temperatures implies that there is not an energy or temperature threshold below which a discrete breather is not excited in a crystal. Work has been published that shows a lower bound in some instances \cite{flach1997,ming2017}, but other work has shown cases in which discrete breathers exist at arbitrarily small energies \cite{kastner2004}. In the case of crystalline solids, the implication from the thermodynamics is that they should fall into the latter rather than the former category.
It is still not clear from the perspective of nonlinear dynamics and numerical studies of models such as the Fermi--Pasta--Ulam--Tsingou system what may happen at long times \cite{carati2007}. Is there thermalization and equipartition at long times, or do these localized excitations persist? Again, the thermodynamics suggests that the latter should be the case.

\section{Discussion}

A possible catch in the foregoing is that as the amplitude approaches that for dissociation the frequency goes to zero and then conditions for existence become hard to satisfy, notably the non-resonance of all harmonics of the breather frequency with that of optical phonons. Nevertheless, one could envisage an aperiodic version of breathers that might play the same role, and indeed include dissociation in its chaotic dynamics without having to wait to gain more energy.
I thank one of the referees for pointing out this caveat.

I am reminded of Wilczek's idea of a \emph{time crystal} in a classical \cite{shapere2012} or quantum system \cite{wilczek2012}, which can be envisaged as an arrangement of atoms that moves in some cyclical manner, so returning periodically to the same initial configuration. 
Watanabe and Oshikawa demonstrated that this behaviour cannot occur in the lowest energy state \cite{watanabe2015}. 
But out of equilibrium, time-crystal-like behaviour has been produced in stable subharmonic oscillations of quantum spin systems \cite{zhang2017,choi2017}.
Meanwhile, in the classical world, 
Yao et al.\ \cite{yao2020} have recently explored classical, discrete time crystals: collective subharmonic oscillations that depend upon an interplay of non-equilibrium driving, many-body interactions and the breakdown of ergodicity.
It seems to me that the persistent localized excitations of the type that we have been interested in here, while not equivalent to the above examples, 
capture the essence of the time crystal idea of a system that is spatially ordered and moves perpetually in an oscillatory fashion. Moreover, the term \emph{time crystal} seems to fit this temporal dynamics of actual, physical crystals.

For solids to be thermodynamically unstable down to 0~K implies that the dynamics even at very low energies must allow the concentration of energy into a single oscillator (a single atom or molecule) for that bond to be able to break. 
The implication of the thermodynamics is that localized excitations such as discrete breathers must exist in general in crystals arbitrarily close to absolute zero.  Nonlinear dynamics then constitutes the means by which the thermodynamical instability of condensed matter in vacuo can be understood.





\bibliographystyle{rsta}
\bibliography{solid_stability}

\end{document}